\documentclass[10pt]{iopart}
\usepackage{iopams}  
\usepackage{graphicx}
\usepackage{epstopdf}
\usepackage[colorlinks,citecolor=blue]{hyperref}
\usepackage{cite}
\begin{document}

\title[$\alpha$-particle densities from ITER CTS]{Inference of $\alpha$-particle density profiles from ITER collective Thomson scattering}

\author{J. Rasmussen$^1$, M. Stejner$^1$, T. Jensen$^1$,  E.~B. Klinkby$^2$, S.~B. Korsholm$^1$, A.~W. Larsen$^1$, F. Leipold$^1$, S.~K. Nielsen$^1$ and M.~Salewski$^1$}

\address{$^1$ Technical University of Denmark,  Department of Physics, Kgs.~Lyngby, Denmark}
\address{$^2$ Technical University of Denmark, Center for Nuclear Technologies, Roskilde, Denmark}

\ead{jeras@fysik.dtu.dk}
\vspace{10pt}

\begin{abstract}
The primary purpose of the collective Thomson scattering (CTS) diagnostic at ITER is to measure the properties of fast-ion populations, in particular those of
fusion-born $\alpha$-particles. Based on the present design of the diagnostic, we compute and fit synthetic CTS spectra for the ITER baseline plasma scenario, including the effects of noise, refraction, multiple fast-ion populations, and uncertainties on nuisance parameters. As part of this, we developed a model for CTS that incorporates spatial effects of frequency-dependent refraction. While such effects will distort the measured ITER CTS spectra, we  demonstrate that the true $\alpha$-particle densities can nevertheless be recovered to within $\sim 10$\% from noisy synthetic spectra, using existing fitting methods that do not take these spatial effects into account. Under realistic operating conditions, we thus find the predicted performance of the ITER CTS system to be consistent with the ITER measurement requirements of a 20\% accuracy on inferred $\alpha$-particle density profiles at 100~ms time resolution.
\end{abstract}

\ioptwocol

\section{Introduction}

In next-step nuclear fusion devices and in future reactors, self-heating of the plasma by fusion-born $\alpha$-particles will be the dominant mechanism for sustaining fusion-relevant conditions. Maintaining good confinement of these $\alpha$-particles will therefore be crucial, and studies of their dynamics in  burning deuterium--tritium (D-T) plasmas thus rank among the main missions of ITER in preparation for future power plants.

In ITER, two diagnostics will be directly sensitive to the $\alpha$-particle velocity distribution function $f_\alpha$, namely $\gamma$-ray spectrometry (e.g.\ \cite{chug11,noce17}) and collective Thomson scattering (CTS) \cite{sale09b,kors16,kors19}. Both will measure $\alpha$-particles at and below their $E\approx 3.5$~MeV birth energy, but $\gamma$-rays will probe these ions only down to $E \approx1.7$~MeV \cite{sale15,sale18}, below which the cross-section becomes negligible for the  most promising $\gamma$--producing reaction, $^9$Be($\alpha$, n$\gamma$)$^{12}$C \cite{kipt90}. Information on the further slowing down of $\alpha$-particles towards their average energy of $E\sim 1$~MeV \cite{walt14} and finally towards thermalization will therefore be provided by CTS alone. As such, CTS will play a major role in diagnosing the velocity distribution and density of confined $\alpha$-particles in ITER.

Such measurements, both for the CTS diagnostic specifically and for ITER in general, are subject to a series of formal measurement requirements. These include the ability to measure the radial density profile of confined fusion-born $\alpha$-particles to an accuracy of 20\% at a time resolution of 100~ms \cite{donn07,idm1,iter}. An integral part of the performance assessment and optimization process for the ITER CTS diagnostic design is hence to evaluate how well the diagnostic can accomplish this for a given proposed CTS mirror geometry.

In this context, we note that analysis of CTS measurements in ITER will differ from that of existing CTS implementations in two important ways. First, the diagnostic 
at ITER will operate in a frequency range of 55--65~GHz in X-mode polarization, below the fundamental electron cyclotron resonance and close to the X-mode L-cutoff  in the plasma \cite{kors16,kors19,rasm19}. This implies that refraction can play a significant role in distorting the measured CTS spectra.
Second, ITER plasmas will contain significant energetic ion populations from both fusion processes and auxiliary heating, with separate ion masses and charges but overlapping velocity ranges. Both of these effects could complicate the interpretation of CTS measurements and require novel analysis considerations, so their impact must be evaluated to obtain a realistic assessment of the diagnostic potential.

Following the initial suggestion to use CTS for diagnosing confined $\alpha$-particles in D-T plasmas  \cite{hutc85}, the proposed design of the ITER CTS diagnostic has evolved considerably from early proposals \cite{orsi97} through feasibility and conceptual design studies \cite{bind03,bind04,meo04,kors06,meo07,tsak08,kors08,mich09a,mich09b} and to the present phase of actual design development, which began in 2014 and is now nearing maturity for the preliminary design review. The development includes the choice of measurement frequencies, the scattering geometries, the number of measurement volumes, and evolution in the ITER machine parameters themselves. Partly based on these evolving diagnostic designs, previous
studies of the potential performance of the ITER CTS diagnostic explored the prospects and nature of both thermal-ion \cite{bind05,kors10,stej12} and fast-ion \cite{eged05,sale09a,sale09b,sale11} measurements. The latter works demonstrated the sensitivity of theoretical ITER CTS spectra to $\alpha$-particles in the presence of additional fast-ion populations from auxiliary heating  \cite{eged05,sale09a,sale09b}. However, this was based on conceptualized measurements relevant to earlier diagnostic designs and did not consider measurable quantities such as anticipated signal and background levels, nor the actual inference of underlying fast-ion properties from such measurements, which had yet to be demonstrated for ITER-like conditions. Hence, it had remained unclear to what extent the diagnostic can actually meet the relevant measurement requirements on $\alpha$-particle densities.

Here we both generate and invert synthetic ITER CTS spectra across the full plasma minor radius, using the actual CTS mirror geometry implemented in the present diagnostic design. The main aim is to establish the accuracy with which spatially resolved $\alpha$-particle densities can be recovered from measurements in the presence of realistic levels of noise, refraction, multiple fast-ion populations, and uncertainties on other plasma parameters. We will also briefly consider the simultaneous recovery of the densities of neutral beam injected fast deuterium ions. Our approach is intended to simulate actual ITER CTS measurements as realistically as possible and to analyze the outcome using currently available methods. This allows the results to be directly compared to the relevant ITER measurement requirements.

The present work includes several novel aspects with respect to earlier studies. First, our analysis is based on the actual front-end design of the CTS diagnostic to be implemented in ITER \cite{kors16,infa17,lope18,vida19,kors19}. We consider results for all seven measurement volumes present in this design, allowing the first reconstruction of full fast-ion density profiles from synthetic ITER CTS spectra. Second, we apply realistic plasma parameters from recent transport simulations of planned ITER operating scenarios, including simulated spatially dependent distribution functions of fusion-born $\alpha$-particles and  neutral-beam injected deuterium. Third, we compute synthetic ITER CTS spectra based on raytracing, taking refractive effects fully into account. To this end, we have developed a novel model for CTS that incorporates spatial effects of frequency-dependent refraction. This determines the location and extent of each measurement volume at a given  measurement frequency, 
while self-consistently accounting for the variation in  plasma and fast-ion parameters covered by each volume at different frequencies.
Fourth, our synthetic spectra incorporate realistic noise levels, based on recent modelling of the electron cyclotron emission from ITER plasmas
\cite{rasm19}, 
and we establish the impact of this on the inference of physical quantities from the spectra. Fifth, we demonstrate for the first time that synthetic ITER CTS spectra can indeed be inverted using existing algorithms to recover spatially resolved information on the underlying fast-ion velocity distribution functions.
The inversion allows for realistic uncertainties in fit priors on nuisance parameters and provides the first demonstration that ITER CTS can meet its measurement
requirements for $\alpha$-particles.

In Section~\ref{sec,methods} we outline our procedure for generating and fitting synthetic ITER CTS spectra. Section~\ref{sec,results} presents the results of trial fits to these spectra under varying assumptions regarding background noise, treatment of nuisance parameters, and modeling of the fast-ion distribution functions. In Section~\ref{sec,discuss}, we discuss these results in the context of the ITER measurement requirements, along with possible caveats and possibilities for improvement. Section~\ref{sec,concl} presents a summary and outlook.

\section{Methods}\label{sec,methods}

The front-end of the ITER CTS diagnostic will include a 1~MW gyrotron beam injecting probing radiation at a frequency $\nu_{\rm gyr} = 60$~GHz, along with seven receiver mirrors 
collecting scattered radiation from separate spatial locations across the frequency range 55--65~GHz  
\cite{kors16,kors19}. The probe beam,
with wave vector ${\bf k}^i$, will interact with (mainly ion-induced) plasma fluctuations with wave vector ${\bf k}^\delta$, to generate scattered waves with wave vector ${\bf k}^s = {\bf k}^i + {\bf k}^\delta$ that can be detected by the  receivers. The scattering spectrum is sensitive to the 1D ion velocity distribution function $g(u)$, which is a projection of the full distribution function $f({\bf v})$ onto ${\bf k}^\delta$, 
\begin{equation}
  g(u) = \int f({\bf v}) \delta \Big(\frac{{\bf v}\cdot {\bf k}^\delta}{k^\delta}-u \Big) \mbox{ d}{\bf v} , 
\label{eq,g_u}
\end{equation}
where $\delta(\ldots)$ is the Dirac $\delta$--function, $u={\bf v} \cdot {\bf k}^\delta/k^\delta \approx 2\pi \Delta \nu/k^\delta$ is the projected velocity along ${\bf k}^\delta$,
and $\Delta \nu = \nu-\nu_{\rm gyr}$ is the frequency shift relative to the probe frequency at which an ion with a given 3D velocity ${\bf v}$ contributes to the spectrum.
Using forward modelling, the measured CTS spectrum can be inverted to recover the underlying $g(u)$. Details of this procedure can be found elsewhere, e.g.\ \cite{bind99,niel08,sale10,niel11,rasm15}.

Each receiver view will intersect the probe beam at a measurement location which depends on the adopted mirror geometry and the plasma refractive index along the receiver viewing path. Schematically, our procedure for generating and fitting synthetic spectra for each receiver comprises the four steps illustrated in Figure~\ref{fig,overview} and summarized as follows:
\begin{itemize}
\item {\em Step 1}: Perform raytracing to get the frequency-dependent scattering geometry for all seven receiver views in the chosen ITER plasma scenario and for the adopted CTS mirror geometry.
\item {\em Step 2}: Compute synthetic ITER CTS spectra: (a) Generate narrow-band spectra with frequency-dependent scattering geometry from {\em Step~1}. (b) Collate to form an aggregate spectrum covering the full 55--65~GHz frequency range. (c) Compute the noise for each resulting frequency bin and perturb the spectrum before fitting.
\item {\em Step 3}: Fit the resulting noisy spectra assuming either exactly known fit priors for nuisance parameters, which are held fixed in the fits, or 
perturbed priors for nuisance parameters, which are treated as free variables in the fits. 
\item {\em Step 4}: Evaluate different measures of fit quality, such as the difference between fitted and true fast-ion densities $n_{\rm fast}^{\rm fit}-n_{\rm fast}^{\rm true}$. 
\end{itemize}
Each step is described in more detail in the following subsections.

\begin{figure}
\begin{center}
\includegraphics[width=6cm]{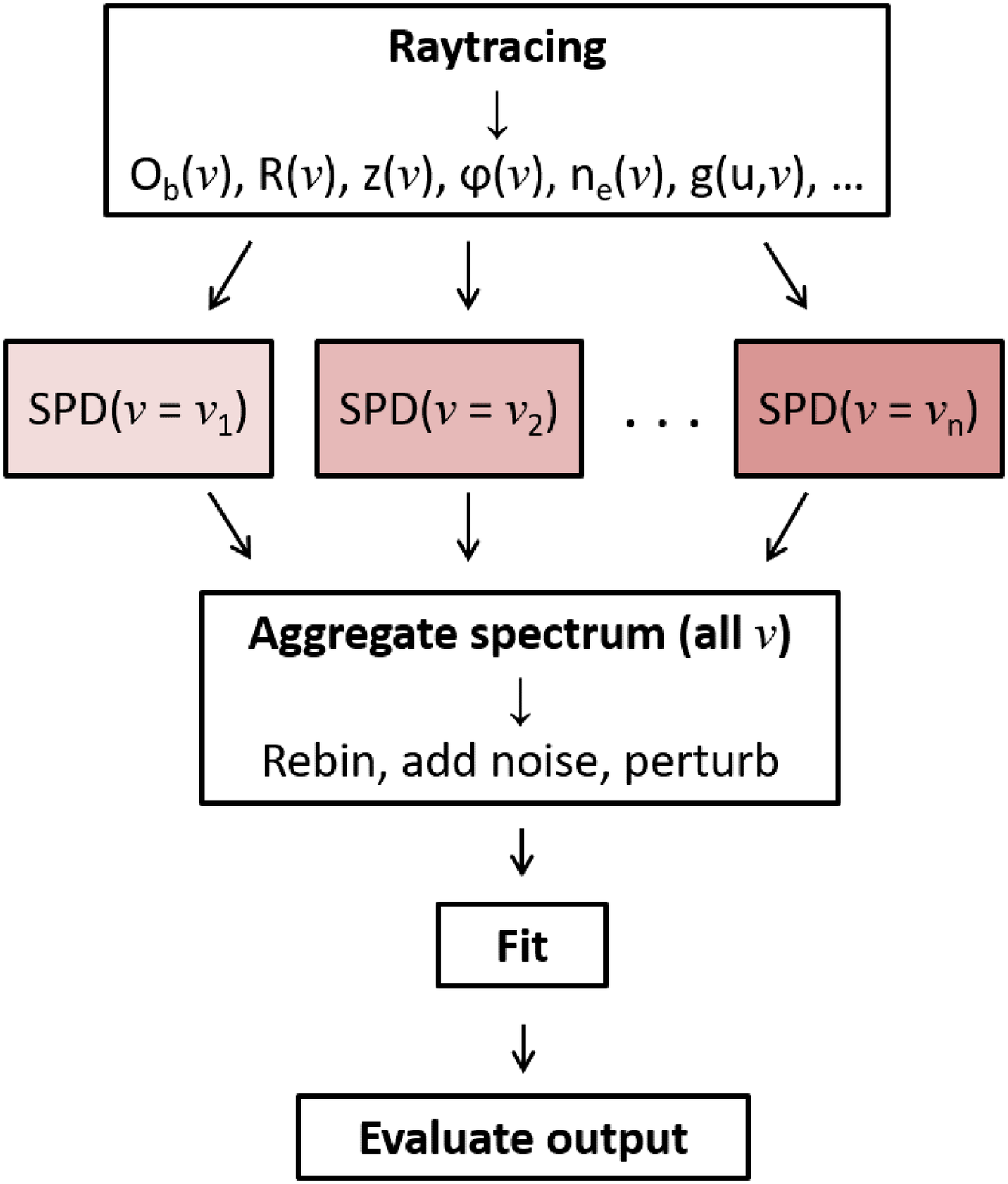}
\caption{Schematic of the fitting procedure. Raytracing done at individual frequencies $\nu$ provides the value of the overlap factor $O_b$ (see Section~\ref{sec,step1}), the location ($R,z$) of the scattering volume, the
scattering geometry $\phi = \angle({\mathbf k^\delta},{\mathbf B})$, and plasma parameters ($n_e$, $g(u)$, \ldots) in the scattering volume. This is used to compute resulting spectral power densities (SPD) at a specific $\nu$, and these are combined into an aggregate spectrum. This spectrum is then manipulated before fitting, and the resulting fit result is finally evaluated.}
\label{fig,overview}
\end{center}
\end{figure}

\subsection{Raytracing}\label{sec,step1}

The presence of refraction has two consequences for the interpretation of CTS data. The most pronounced one is that the region  sampled by a given CTS receiver in real space and fast-ion velocity space can differ dramatically from the vacuum expectation. For existing CTS diagnostics, this is usually dealt with in post-experiment analysis by reconstructing the experimental scattering geometry using raytracing. We assume here that this can be done for the diagnostic at ITER in a similar manner and with comparable uncertainties. A more subtle effect is that, due to the frequency dependence of refraction, the properties of the scattering volume, from which the CTS signal originates, can vary somewhat with frequency.
In that case, the signal at each measured frequency arises under a separate scattering geometry and so samples a specific region in real space and fast-ion velocity space. 
Both of these effects will be illustrated below when discussing Figures~\ref{fig,profiles} and \ref{fig,raytracing}.

To account for this, we use our raytracing code {\em Warmray}, developed by H.~Bindslev. This code evolves ray trajectories  in the WKB approximation and has been successfully used in the design and interpretation of  numerous CTS experiments in past and present devices  (e.g.\ \cite{niel17}). It  includes effects of relativistic electrons on the plasma refractive index relevant for ITER plasmas \cite{bind92a,bind92b,bind93}, based on the weakly relativistic dielectric tensor and plasma susceptibilities of \cite{shka86}. Here we apply the code at individual frequencies spaced 10~MHz apart, to provide the frequency-dependent viewing path of each receiver and the associated scattering geometry. The latter is partly defined by the
projection angle $\phi = \angle({\mathbf k^\delta},{\mathbf B})$, and by the
 location and value $O_b$ of the beam overlap \cite{bind91}, 
\begin{equation}
O_b = \int \mathcal{I}_i(\mathbf{r},t) \mathcal{I}_s(\mathbf{r},t) \mbox{ }d\mathbf{r},
\end{equation}
which quantifies the degree to which the probe and receiver beams intersect and hence determines the overall scaling of the measured CTS spectrum. Here
$\mathcal{I}_i$ and $\mathcal{I}_s$ are the normalized intensity distributions of the incident and received scattered radiation, respectively.

For the plasma parameters used in raytracing, we consider the ITER baseline ($B_t = 5.3$~T, $I_p = 15$~MA) H-mode scenario, which is expected to comprise the
majority of D-T pulses and whose establishment is the first major goal of the ITER Research Plan 
\cite{hend15,iter18}. In order to emphasize the impact of refraction, we have adopted electron density and temperature profiles
from JINTRAC simulations \cite{scen1} that predict a slightly higher central $n_{e,0} \approx 12.5 \times 10^{20}$~m$^{-3}$ and lower central $T_{e,0} \approx 17$~keV than suggested by several other simulations of the ITER baseline scenario, including some of
those incorporated in the scenario database of the ITER Integrated Modelling \& Analysis Suite (IMAS; \cite{imbe15,pinc18}). This allows us to assess the  impact of refraction on the interpretation of CTS spectra under potentially conservative conditions. 

Figure~\ref{fig,profiles} outlines the resulting plasma profiles, described in more detail in Section~\ref{sec,step2}.  As required by {\em Warmray}, profiles of $n_e$ and $T_e$ have been extrapolated beyond $\rho_p=1$, here using a modified {\tt tanh} function \cite{schn12} and a smoothed linear decline, respectively. Figure~\ref{fig,raytracing} shows the scattering geometry for all seven ITER CTS receivers in this plasma scenario at frequencies $\pm 4$~GHz away from the 60~GHz probe frequency, demonstrating the impact of refraction itself as well as its frequency dependence. Each receiver is labelled in the following according to the location of the corresponding scattering volume, as illustrated in  Figure~\ref{fig,raytracing}. Note that the measurement volume of Receiver~1 is located on the high-field side of the plasma, and that all receiver views, in particular those close to the plasma edge, are subject to significant refraction due to the proximity of the X-mode L-cutoff \cite{rasm19}. Refraction also acts in the toroidal direction, but this effect is similar for the probe and receiver ray paths, thus maintaining the overlap factors.

\begin{figure}
\begin{center}
\includegraphics[width=8cm]{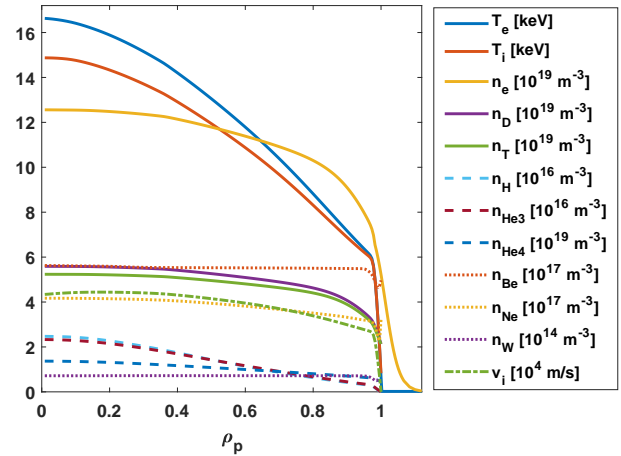}
\caption{Plasma profiles in the adopted ITER baseline plasma scenario as functions of poloidal flux coordinate, showing electron and ion temperatures $T$, densities $n$ of the various thermalized plasma species, and the ion toroidal rotation velocity $v_i$.}
\label{fig,profiles}
\end{center}
\end{figure}

\begin{figure*}
\begin{center}
\vspace{8mm}
\includegraphics[width=5cm]{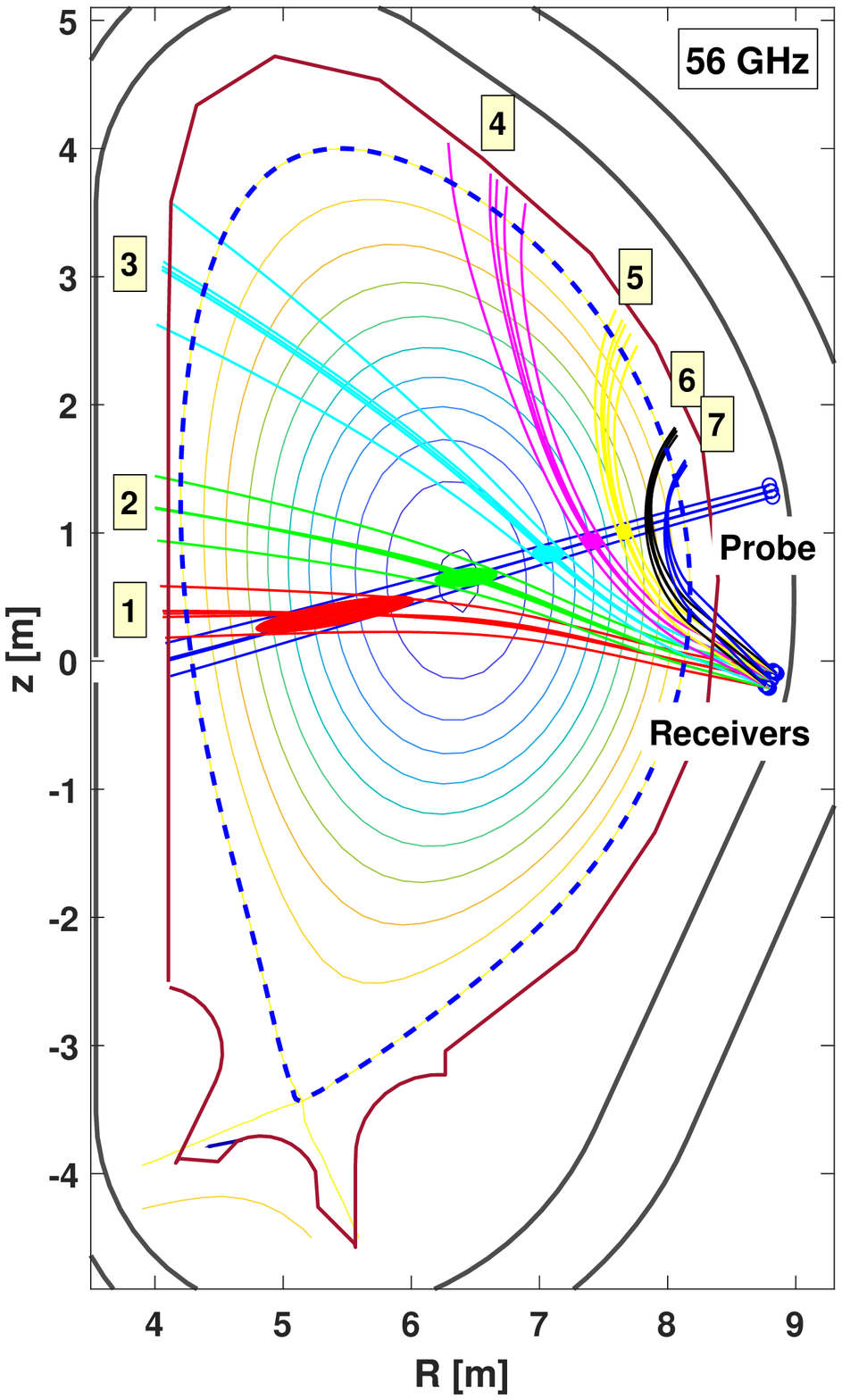}\hspace{5mm}
\includegraphics[width=5cm]{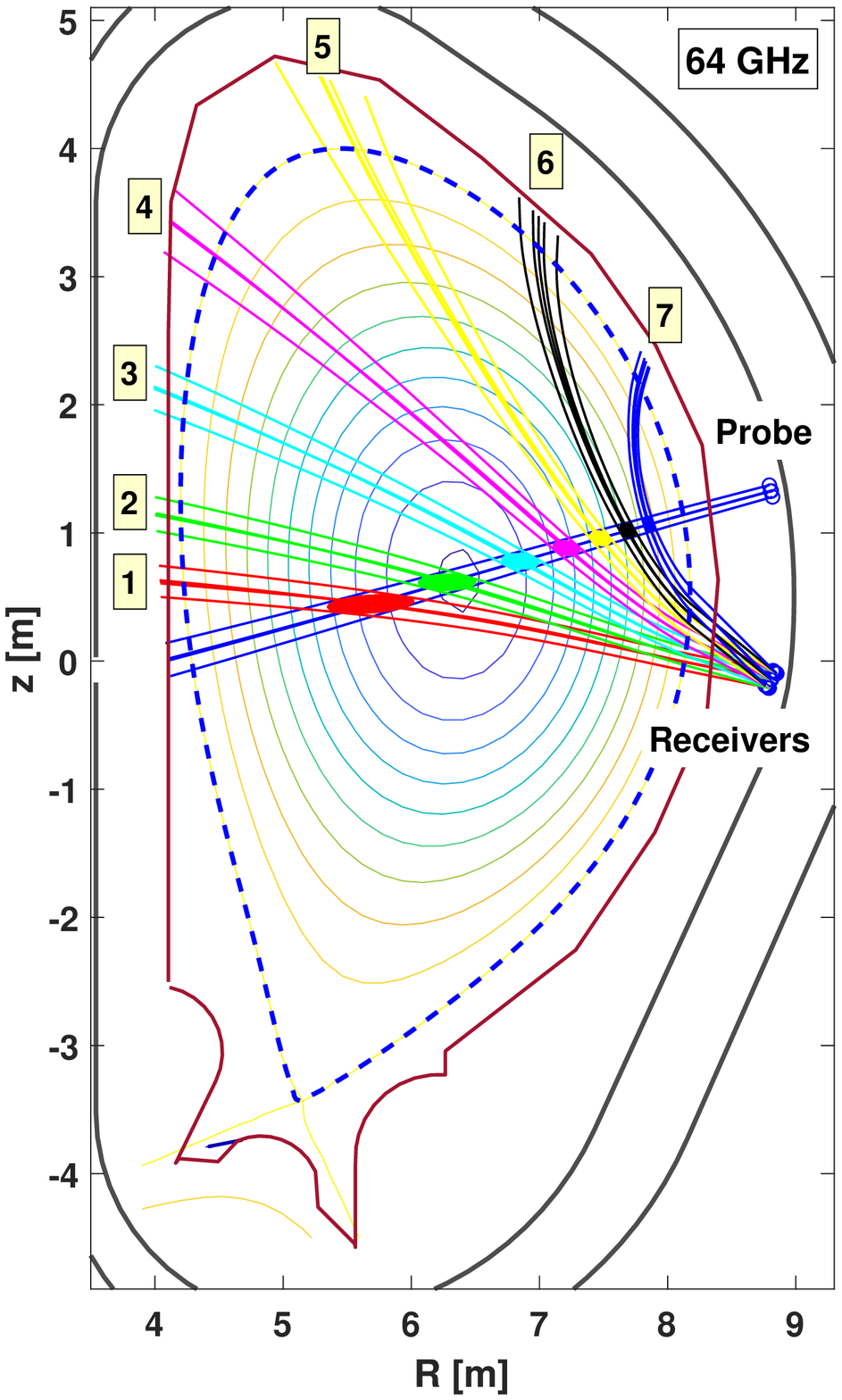}
\caption{Scattering geometries of the seven ITER CTS receivers in a poloidal cross section, based on raytracing at 56 and 64~GHz in the adopted baseline plasma scenario (probe
beam is at 60~GHz). Ellipses outline the extent of the scattering volumes, here defined as the region containing 75\% of the scattering signal. Receivers are (arbitrarily) labelled in order of increasing major radius $R$ of the corresponding scattering volume.}
\label{fig,raytracing}
\end{center}
\end{figure*}

\subsection{Computing and perturbing synthetic CTS spectra}\label{sec,step2}

Given the scattering geometries resulting from {\em Step~1}, we next calculate synthetic CTS spectra for a given scattering geometry and plasma location including contributions from all thermal and fast ions.  This is done using a fully electromagnetic model of the scattering process \cite{bind93b,bind96}. 
The synthetic spectra consist of contributions from thermal (Maxwellian) electrons, D and T \cite{scen1}, the fusion products H, $^3$He, and $^4$He \cite{fast}, and the impurities Be, Ne, and W \cite{scen1}. All are subject to toroidal rotation velocities adopted from \cite{scen1}, and their radial profiles are outlined in Figure~\ref{fig,profiles}.
In addition, simulated fast-ion distribution functions for fusion-born $^4$He and for neutral beam--injected D were included from \cite{fast}. 

To account for frequency-dependent refraction, the spectral power densities of the scattered waves are calculated for each frequency considered in the previous raytracing step. These power densities are then combined into an aggregate spectrum covering the full frequency range. This accounts for the fact that the signal at each frequency is generated at a specific location in the plasma with a specific scattering geometry and local plasma and fast-ion parameters. Resulting spectra for selected receivers will be presented in Section~\ref{sec,refrac}.

The model spectra, computed at a frequency resolution of 10~MHz corresponding to 1,000 frequency bins, are then prepared for fitting. This consists of re-binning the spectra and incorporating noise estimates. The spectra are partitioned into a desired lower number of frequency channels by averaging the initial model spectrum across the frequencies of each channel. This is to emulate the limited number of frequency channels typical of present-day receiver setups (using filter banks for fast-ion measurements), increase the signal-to-noise ratio of the final synthetic spectra (see below), and to reduce the computational load associated with fitting. 

In order to include uncertainties on all data points in the resulting spectra, we consider three separate noise contributions. The first is the usual standard deviation of the estimate of the post-integration CTS signal \cite{bind92b}, i.e. 
\begin{equation}
\sigma_{\rm CTS} = \left(\frac{2(P_s+P_b)^2+2P_b^2}{W \Delta t}\right)^{1/2},
\end{equation}
where $P_s$ and $P_b$ is the spectral power density of the CTS signal and electron cyclotron emission (ECE) background, respectively, falling within a specific frequency channel of width $W$, and $\Delta t$ is the total useful integration time (gyrotron on+off period). This expression has been shown to provide a reliable estimate of the noise in bulk-ion (i.e.\ high frequency resolution) CTS measurements at both TEXTOR and ASDEX Upgrade \cite{stej10,stej17b}. 

The ITER measurement requirements are defined for an integration time of $\Delta t = 100$~ms. Here we will as default split this into five shorter segments of $\Delta t = 20$~ms, allowing us to average the fit results pertaining to those individual segments and derive a statistical uncertainty, and hence characteristic accuracy, on the results for a 100~ms measurement period.
For $P_b$, detailed ECE modelling \cite{rasm19,stej18} suggests that the ITER ECE background in the baseline scenario should not exceed $\sim 10$~eV at the relevant frequencies for realistic reflectivities of the first wall. Here we take $P_b = 100$~eV across all frequencies as a deliberately conservative choice. Finally, the receiver channel width $W$ is set by the need to resolve the spectrum out to $|\nu - \nu_{\rm gyr}| = 5$~GHz for the chosen number of frequency channels.

The above noise estimate assumes that $P_s$ varies negligibly across the bandwidth $W$. To account for any large variations in $P_s$ across the 
adopted frequency channels, a noise contribution $\sigma_{\rm sd}$ equal to the standard deviation of $P_s$ within each channel is also included. This term, derived from the unbinned spectra from {\em Step~2} above, dominates the noise at frequencies where the spectrum is very steep, i.e.\ in the vicinity of the thermal bulk and any magnetosonic wave features. Finally, an estimated systematic uncertainty is included of $\sigma_{\rm sys} = 0.5$~eV from, e.g., background subtraction uncertainties and noise in the receiver transmission line and electronics. This estimate is based on experience from the CTS diagnostic at ASDEX Upgrade \cite{rasm15,niel15,rasm16} and dominates the noise at low $P_s$ and $P_b$. 

Assuming that the various noise contributions are uncorrelated, the total noise estimate $\sigma_{\rm tot}$ for each frequency channel is then obtained as 
\begin{equation}
\sigma_{\rm tot}^2 = \sigma_{\rm CTS}^2 + \sigma_{\rm sd}^2 + \sigma_{\rm sys}^2.
\end{equation}
With these assumptions, and taking $W = 200$~MHz corresponding to 50 frequency channels (comparable to the CTS filter bank setup at ASDEX Upgrade \cite{furt12}), the final uncertainties are generally not dominated by the contribution from $P_b$. This applies unless $P_b \gtrsim 0.5$~keV at these frequencies, which would require background levels far exceeding theoretical expectations for approximately Maxwellian electron populations in the baseline plasma scenario \cite{rasm19}.
Based on this combined noise estimate, each data point in the spectrum is finally perturbed within its (assumed Gaussian) uncertainty $\sigma_{\rm tot}$ before fitting. Figure~\ref{fig,spec}({\em a}) shows an example of the original aggregate spectrum and the final rebinned and perturbed spectrum generated in this manner for our assumed integration time of $\Delta t= 20$~ms.

\begin{figure}
\begin{center}
\includegraphics[width=8.4cm]{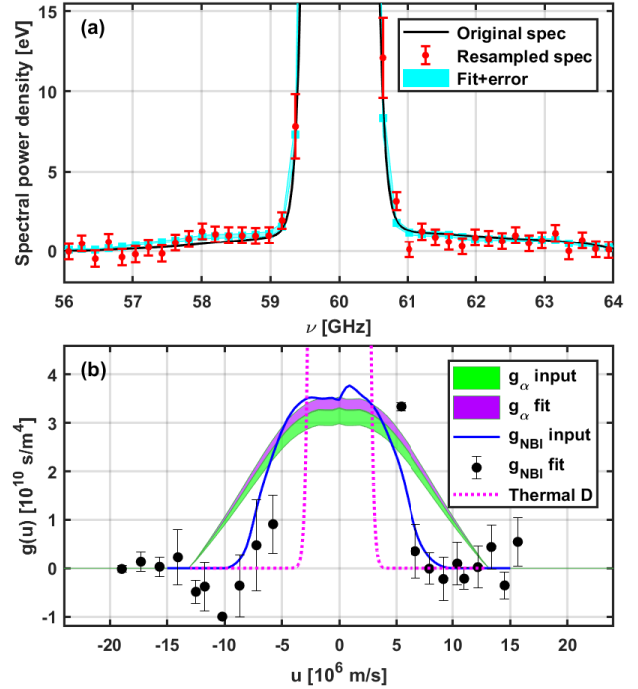}
\caption{({\em a}) Original model spectrum (black curve) for Receiver~1, along with the final spectrum to be fitted based on rebinning and randomizing the original spectrum according to our default noise estimates for a $\Delta t=20$~ms acquisition period. Cyan curve shows the spectrum and associated uncertainties corresponding to the best-fit forward model.
({\em b}) 1D distribution functions $g(u)$ for $\alpha$-particles and fast-D ions underlying the spectrum shown in ({\em a}), along with the corresponding average fitted distributions and their errors obtained from five such individual spectra (corresponding to $\Delta t=100$~ms). The distribution function for thermal D, the lighter of the two fuel-ion species, is shown for comparison.}
\label{fig,spec}
\end{center}
\end{figure}

\subsection{Fitting the spectra}\label{sec,step3}

The fitting is performed within a Bayesian framework \cite{bind99,stej12}, using Bayesian priors on the scattering geometry and on plasma parameters in the measurement volumes obtained from raytracing at the probe frequency. Effectively, this means that the varying dependence of refraction across the fitted frequency range is ignored in the data analysis, as it is intractable to correct for this for every 20~ms data acquisition period. Our approach thus allows us to assess the impact of neglecting this effect when inverting the measurements.
 
The focus here is on the fast-ion densities derived from integrating the fitted $g(u)$ from equation~(\ref{eq,g_u}) over an observable projected velocity interval. Here,
$g(u)$  is modelled as a free, non-parametric distribution in our fitting algorithm, where it represents a single distribution function for a specific fast-ion species. Hence, a complication resulting from the presence of  both fusion $\alpha$-particles and beam-injected D in our synthetic data, is that the fitted $g(u)$ must simultaneously account for all fast ions in the plasma, but it is not possible to differentiate between different species by CTS. For example, an $\alpha$-particle and a D ion at the same projected velocity $u$ cause scattering in the same frequency range with the same functional form. The amplitude of the scattering signal is proportional to the square of the ion charge \cite{hugh88}, and hence the signatures of an $\alpha$-particle and four D ions with 
the same $u$ are indistinguishable.

However, the fusion $\alpha$-particles can be assumed to be distributed according to a classical, isotropic slowing-down distribution to a fairly good approximation. Such a model
 is implemented here as an optional fit component  in addition to $g(u)$, which is then assumed to describe only the fast D ions from NBI.
The slowing-down model contains three free parameters, namely the slowing-down critical velocity $v_c$ (at which electron and ion collisions contribute equally to the $\alpha$ slowing-down), the $\alpha$-particle birth energy, and the overall scaling.  For $v_c$, which depends on $T_e$ and very weakly on the $n_D/n_T$ ratio, we assume a relative uncertainty corresponding to that on $T_e$, while the $\alpha$-particle birth energy of $E = 3.52$~MeV is assumed fixed with no uncertainty. The Bayesian prior on the scaling, i.e.\ on the total $\alpha$-particle density $n_\alpha$, is assumed to be highly uncertain, as described in Section~\ref{sec,results}. In a real experiment, the $\alpha$-particles will have small deviations from a classical slowing-down distribution. We simulate this here by using a numerical simulation of $f_\alpha$ when computing the synthetic spectra.

We are free to chose the number of velocity nodes when discretizing the  reconstructed $g(u)$. For our default choice of 50~frequency channels in the final synthetic spectra,
we here adopt 20 velocity nodes outside the thermal bulk. The impact of modifying the number of velocity nodes and spectral frequency channels will be briefly discussed in Section~\ref{sec,discuss}. Since the perturbed spectra to be fitted may contain signal levels that are nominally negative at large frequency shifts due to the background noise (cf.\ Figure~\ref{fig,spec}({\em a})), so can individual velocity nodes in the reconstructed $g(u)$. We do not enforce any non-negativity constraint on these individual nodes.

Figure~\ref{fig,spec}({\em b}) illustrates the overall approach, for a case with the $\alpha$-particles treated by a slowing-down distribution, $g(u)$ representing fast D from neutral beam injection (NBI), and where the peak velocity-space densities of the two populations are comparable. The figure shows the simulated $\alpha$-particle distribution function (green curve) -- with uncertainties as described in Section~\ref{sec,step4} -- as well as the corresponding fitted slowing-down distribution including fit uncertainties (purple curve). Also shown are the true and fitted $g(u)$, along with the distribution function of bulk D ions in order to illustrate the region at small projected velocities $|u|$, in which the fast-ion signal is masked by the thermal contribution. The figure represents the average of five fits to synthetic spectra, each representing our default integration time $\Delta t = 20$~ms, such that the fit results correspond to $\Delta t = 100$~ms, the time resolution of the relevant ITER measurement requirements. A forward model based on one of those fits, that corresponding to the spectrum in Figure~\ref{fig,spec}({\em a}), is overlayed on the latter figure. Note that while the particular realization of fit results based on a 100~ms integration time shown in Figure~\ref{fig,spec}({\em b}) seems to slightly overestimate the $\alpha$-particle contribution, the average of many such results does not show any systematic offset (see Section~\ref{sec,series3}).

In addition to the fast-ion distribution function, CTS spectra depend on a range of other plasma parameters that are of secondary interest for the present purposes. In most of the fit cases considered below, Bayesian priors on these nuisance parameters in the fit are also randomly perturbed within their assumed Gaussian uncertainties. Where not indicated otherwise, these uncertainties, summarized in Table~\ref{tab,nuisance}, correspond to the target uncertainties from the ITER diagnostics database \cite{donn07,idm1,iter}. 
 Finally, we assume an injected gyrotron power of $P_{\rm gyr}=1.0$~MW in both model spectra and fitting, with no uncertainty. 

\begin{table}
\caption{\label{tab,nuisance} Uncertainties on Bayesian priors for nuisance parameters in fits to synthetic spectra.} 
\footnotesize
\begin{tabular}{@{}ll}
\br                              
Parameter & $\sigma_{\rm prior}$ \cr
\mr
Absolute scaling & 20\%$^{a}$ \cr            
Scattering angles & $1^\circ$$^a$ \cr
Magnetic field strength $B$ &	0.1 T$^{b}$\cr
Fuel ion ratio $R_i = n_T/(n_D+n_T)$ &	24\%$^{c}$ \cr
Ion temperature $T_i$ &	10\% \cr
Ion drift velocity $v_i$ &	20\% \cr
Hydrogen density $n_H$ &	14\%$^{c}$ \cr
Thermal He densities $n_{\rm He3}$,  $n_{\rm He4}$ &	20\%$^{d}$ \cr
Other impurity densities (Be, Ne, W) & 10\% \cr
Electron density $n_e$ &	5\% \cr
Electron temperature $T_e$ & 	1\% \cr
\br
\end{tabular}\\
$^{a}$Assumed uncertainty on the overall scaling (representing uncertainty in the overlap factor) and scattering geometry as reconstructed from raytracing. $^{b}$Expected measurement uncertainty on $B_t$ is only 0.3\%, but here we take $\sigma_B = 0.1$~T to allow for raytracing uncertainties on the location of the scattering volume and hence on the local magnetic field. $^{c}$The relative target uncertainty $\sigma$ is 0.2 on both $n_T/n_D$ and $n_H/n_D$. If assuming $\sigma_D = \sigma_T= \sigma_H$, then $\sigma_H/n_H=0.14$ and $\sigma_{R_i}/R_i =0.24$. $^{d}$Unlike for other species, the target uncertainty on thermal $n_{\rm He}$ is defined relative to $n_e$ \cite{iter}, with $\sigma_ {\rm He/e} \lesssim 5$\% at flux coordinates $\rho \gtrsim 0.3$. Assuming $^4$He only, this would correspond roughly to $\sigma_{\rm He4}/n_{\rm He4} \lesssim 0.3$ for the profiles in Figure~\ref{fig,profiles}. Here we take $\sigma=20$\% for $^3$He and $^4$He individually relative to their absolute densities.

\end{table}
\normalsize

\subsection{Evaluating the quality of the fits}\label{sec,step4}

For the comparison of fit results to true parameters, we use the synthetic $g(u)$ computed on the basis of raytracing at the gyrotron frequency only. The focus here is primarily on the density $n_\alpha$ of $\alpha$-particles and secondarily on that of fast D from NBI injection, $n_{\rm NBI}$, in the projected velocity interval. We ignore for now that the primary measurement targets for the ITER CTS diagnostic include also fast H, T, and $^3$He from D-D fusion, as predicted fast-ion distribution functions for these species are not currently available to us for any ITER scenario. We expect the determination of $n_\alpha$ to be fairly unaffected by the presence of these additional species in realistic ITER D-T plasmas, since $\alpha$-particles will still dominate the CTS signal at large frequency shifts due to the higher cross section of D-T fusion, the higher birth energy of $\alpha$-particles, and their larger charge compared to fast H and T.
 
As a first step, the results are evaluated in terms of how well the true velocity-space density $n_{\rm fast}$ of fast ions can be recovered. In terms of addressing the ITER CTS measurement requirements, the accuracy and uncertainty of the fitted $n_{\rm fast}$  provide an intuitive and straightforward measure of diagnostic performance. When $\alpha$-particles are accounted for in the fit by a separate slowing-down distribution, the total fitted $\alpha$-particle density can be directly compared to an integral over the true $\alpha$-particle distribution function. When plotting the true $\alpha$-particle density profiles derived in this way, we will include a 5\% uncertainty when relevant, since
the total ion density associated with a slowing-down distribution in our fitting code is provided relative to $n_e$, whose Bayesian prior can vary according to Table~\ref{tab,nuisance}.

For $g(u)$, the fit results are evaluated in terms of {\em observable} (or "partial") fast-ion densities integrated over the projected velocity range outside the thermal bulk, as defined by the location of the velocity nodes in the fit. The result can be compared to the similarly computed true density, but the latter will generally differ significantly from the {\em total} true density, due to the exclusion of velocity nodes around the thermal bulk. Indeed, since the projection angle $\phi = \angle({\mathbf k^\delta},{\mathbf B})$ is close to perpendicular to the magnetic field  for the adopted CTS mirror geometry (see next section), a large fraction of fast ions are masked by the thermal bulk ions in the resulting CTS spectra. Their observable densities outside the thermal bulk are therefore considerably lower than the true total values. As an example, the true total NBI density corresponding to the blue curve in Figure~\ref{fig,spec}({\em b}) of $n_{\rm NBI} = 4.3\times 10^{17}$~m$^{-3}$ is a factor of $\sim$\,8  larger than the corresponding observable fast-ion density of $0.5 \pm 0.1 \times 10^{17}$~m$^{-3}$ obtained by integrating $g(u)$ only outside the range defined by the innermost velocity nodes (black data points).

\section{Results}\label{sec,results}

\subsection{Impact of frequency-dependent refraction}\label{sec,refrac}

Before addressing the fit results, we first consider the effect of refraction on the modelled spectra. Figure~\ref{fig,refrac} shows example results which compare spectra computed with frequency-dependent refraction to those produced in the standard fashion based on raytracing at the probe frequency alone. For the adopted diagnostic mirror geometry and plasma scenario, the projection angle at $\nu_{\rm gyr} = 60$~GHz ranges between $\phi = 96^\circ$--$100^\circ$ for the seven receivers and so is relatively close to perpendicular to the magnetic field. This enables the fast magnetosonic wave to affect the spectra at the frequency shift which satisfies its dispersion relation \cite{bind96}, producing a clear spectral peak at large $\Delta \nu$ for some of the receivers.
However, as will be shown, this does not significantly hamper the recovery of fast-ion information from the spectra.

\begin{figure*}
\begin{center}
\includegraphics[width=17cm]{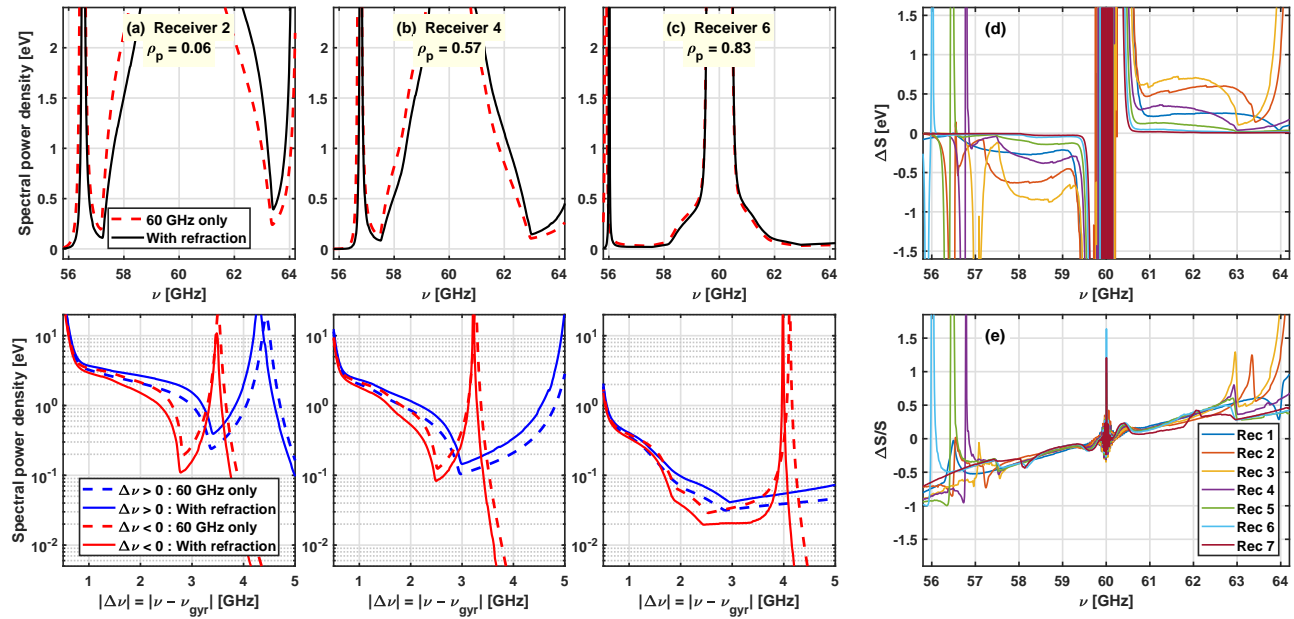}
\caption{Example ITER CTS spectra computed with and without frequency-dependent refraction for ({\em a}) Receiver~2, ({\em b}) Receiver~4, and ({\em c}) Receiver~6, labelled with $\rho_p$ of the central measurement location at 60~GHz. These are shown on a (top) linear scale as a function of absolute frequency, and (bottom) logarithmic scale and relative to the probe gyrotron frequency $\nu_{\rm gyr} = 60$~GHz. Panel ({\em d}) shows for all receivers the absolute difference $\Delta S = S_{\rm refrac} - S_{\rm 60 GHz}$ between spectra computed with full refractive effects and those computed at $\nu_{\rm gyr}$  only,  and ({\em e}) shows the relative difference $\Delta S/ S_{\rm 60 GHz}$.} 
\label{fig,refrac}
\end{center}
\end{figure*}

The frequency-dependent impact of refraction is generally modest, even in this high-density plasma scenario. The most important effect is to slightly distort the spectra, reducing the signal at downshifted frequencies and increasing it at upshifted ones. Lower frequencies closer to the X-mode L-cutoff suffer stronger refraction, so these generally intersect the probe beam further off-axis compared to the expectation at the probe frequency itself, thus reducing the scattered power. The opposite is generally true at higher frequencies. 
While typical relative differences induced by this effect may exceed 50\% at the large frequency shifts $\Delta \nu \propto {\mathbf v_{ion}} \cdot  {\mathbf k}^\delta$ relative to $\nu_{\rm gyr}$ that are relevant to fast ions, the absolute changes typically amount to no more than $\sim 0.5$~eV outside the magnetosonic peaks in the spectra. In most cases, this is within our adopted systematic noise floor of $\sigma_{\rm sys}= 0.5$~eV. The resulting variation $\Delta \phi$ in the projection angle between 55 and 65~GHz is also small, $\Delta \phi \lesssim 1^\circ$ for all receivers.

\begin{table*}
\caption{\label{tab,fits} Summary of fit procedures.} 
\footnotesize
\begin{tabular}{@{}*{6}{c}}
\br             
            Fit series &	Noise level &	Nuisance &	$\alpha$'s fitted with & $\alpha$-particle & Subsection \cr
 & & parameters & slowing-down & density prior & \cr
                          &         &  fixed (Y/N)                                                 &    distribution (Y/N) & & \cr
\mr
1 &	Reduced &	Y & Y &	$10^{14}\pm 5 \times 10^{18}$~m$^{-3}$ & \ref{sec,series1} \cr 
2 &	Nominal &	N &	Y&	$10^{14}\pm 5 \times 10^{18}$~m$^{-3}$ & \ref{sec,series3} \cr 
3 &	Enhanced &	N  &	Y &	$10^{14}\pm 5 \times 10^{18}$~m$^{-3}$ & \ref{sec,series4} \cr 
4 &	Nominal & N	   &  N	& None &	\ref{sec,series5} \cr 
\br
\end{tabular}
\end{table*}
\normalsize

Turning now to the fit experiments, different types of fits were performed to test the recovery of $\alpha$-particle  and fast-ion (NBI or NBI + $\alpha$) densities under a range of assumptions, as summarized in Table~\ref{tab,fits} and described in detail below.

\subsection{Fit series 1: Reduced noise and fixed nuisance parameters}\label{sec,series1}

We will first consider a case that exposes only the effect of frequency-dependent refraction and the degenerate functional dependence on $n_\alpha$ and $n_{\rm NBI}$. Here, all nuisance parameters except the overall scaling are fixed in the fit, and their Bayesian priors remain unperturbed. However, uncertainties on these parameters according to Table~\ref{tab,nuisance} are still taken into account in fitting. This is equivalent to assuming that plasma conditions can be determined from other ITER diagnostics with sufficient accuracy that the relevant parameters need not be fitted along with the fast-ion distribution functions when inverting the CTS measurements.

In addition, we set $P_b = 1$~eV and  $\sigma_{\rm sys} = \sigma_{\rm sd} = 0$. The motivation for this is partly to remove these sources of uncertainty in this initial test case, and partly that the ECE background at the relevant frequencies might well be negligible \cite{rasm19}, while systematic noise contributions can potentially be brought under control and  suppressed by a suitable choice of frequency channels. By minimizing these sources of noise, we can isolate any discrepancies between the fitted and true fast-ion densities to three main effects, namely frequency-dependent refraction, degeneracy of the functional dependence on $n_\alpha$ and $n_{\rm NBI}$, and using an analytical model to fit the simulated $\alpha$-particle contribution.  The discrepancies due to these effects will be shown to be small.

The fast-ion velocity distribution is here represented by the two components discussed in Section~\ref{sec,step3}: (a) An $\alpha$-particle component assumed to have a classical slowing-down distribution; and (b) a fast deuterium component represented by a non-parametric $g(u)$ discretized into 20~velocity nodes.
The fits assume an initially "unknown" $\alpha$-particle density by using an almost flat Gaussian fit prior ($n_\alpha = 10^{14}$~m$^{-3}$, $\sigma=5 \times 10^{18}$~m$^{-3}$) in the relevant data range ($5 \times 10^{15}$ -- $9 \times 10^{17}$~m$^{-3}$) probed by any of the receivers. Similarly, we assume no prior knowledge of $n_{\rm NBI}$.

Figure~\ref{fig,series1} shows examples of results produced under these assumptions. As explained in Section~\ref{sec,step2}, the profiles shown here were obtained by averaging results of five individual trial fits to five individual synthetic spectra generated for each receiver. The results thus represent a combined integration time of $\Delta t = 5 \times 20$~ms = 100~ms, 
and the plotted error bars represent the standard error on the mean of these five trial fits. This plot, along with all similar subsequent ones (unless specified otherwise), is thus intended to show a {\em representative} example of a reconstructed density profile and the associated uncertainties for a 100~ms measurement,
corresponding to the time resolution of the ITER measurement requirements. Figure~\ref{fig,series1}({\em a}) shows the true $\alpha$-particle density used to calculate the synthetic spectra, the fit prior on $n_\alpha$ in the fits, and the $\alpha$-particle density inferred from the fits, as well as the relative discrepancy between the fitted and true values, all as a function of $\rho_p$ of the central measurement location for each receiver. Similarly, Figure~\ref{fig,series1}({\em b}) shows the observable true and fitted NBI fast-ion densities obtained as described in Section~\ref{sec,step4}. 

For this case, the true $\alpha$-particle densities are very well reproduced, with the density for all individual receivers recovered to within $0.3 \times 10^{17}$~m$^{-3}$ and with an average absolute offset from the true values of $0.15\pm 0.04 \times 10^{17}$~m$^{-3}$  (mean and $1\sigma$ error for all receivers across five trial fits). For the adopted CTS mirror geometry and plasma parameters, the innermost five receivers at $\rho_p < 0.8$ probe plasma locations corresponding to $n_\alpha \gtrsim 1\times 10^{17}$~m$^{-3}$, which is the lower density limit set in the ITER measurement requirements for CTS. Defining a single figure of merit as the relative 
accuracy in recovered $\alpha$-particle densities averaged across these five receivers, i.e.\
\begin{equation}
\Delta^*_{\rm rel} = \Big \langle \frac{|\langle n_{\alpha,{\rm fit}}\rangle -n_{\alpha,{\rm true}}|}{n_{\alpha,{\rm true}}}  \Big \rangle_{\rm Rec 1-5}  ,
\label{eq,drel}
\end{equation}
 we find $\Delta^*_{\rm rel} = 0.07\pm 0.03$ in this case  (mean and $1\sigma$ error), i.e.\ $n_\alpha$ is here on average recovered to within $\approx 7$\% above this 
$\alpha$-particle density limit.

The fitted observable fast-D densities are generally lower, but also comparable to the corresponding true values. This shows that the fitting algorithm correctly separates the degenerate overall contributions of $\alpha$-- and NBI particles to the spectra, despite the absence of useful prior constraints on the total density of either species.
The slight kink in the true observable fast-ion profile seen in the figure will be discussed in more detail in Section~\ref{sec,discuss}. It is a consequence of differences in scattering geometry among the receiver views along with the way velocity nodes are distributed by the fitting algorithm when reconstructing $g(u)$.

\begin{figure}
\begin{center}
\includegraphics[width=8.2cm]{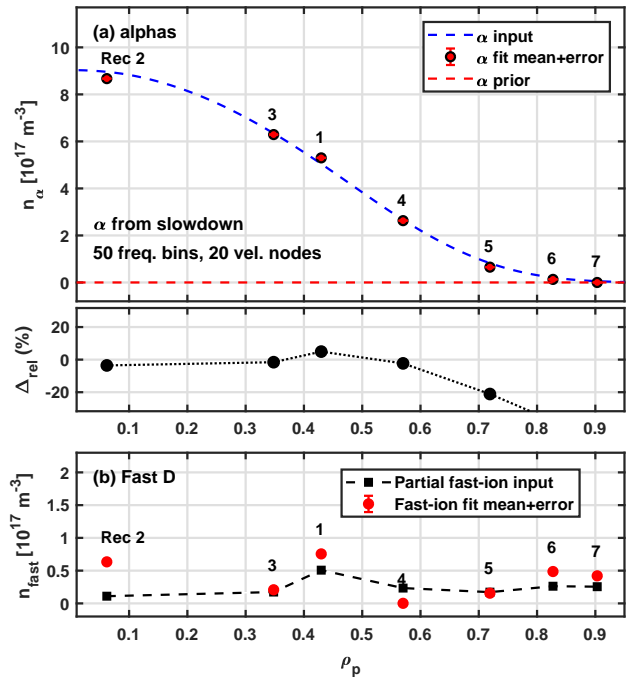}
\caption{Example results of fit series~1 with reduced noise and fixed nuisance parameters. ({\em a}) Results for the $\alpha$-particle densities integrated over all projected velocities, along with the residuals relative to the true values, $\Delta_{\rm rel} =  (n_{\rm fit}-n_{\rm true})/n_{\rm true}$. ({\em b}) Results for the total and partial (i.e.\ observable, excluding the thermal bulk) NBI fast-ion densities. $1\sigma$ error bars on data points are plotted but are not visible. Central measurement locations of the CTS receivers at 60~GHz are labelled according to Figure~\ref{fig,raytracing}.} 
\label{fig,series1}
\end{center}
\end{figure}

\subsection{Fit series 2: Nominal noise and free nuisance parameters}\label{sec,series3}

We now allow for nominal noise levels, as well as discrepancies between the true and Bayesian prior values for nuisance parameters related to the electrons and thermal ions. Hence, the Bayesian priors for all nuisance parameters fluctuate within the uncertainties in Table~\ref{tab,nuisance}, and the parameters themselves (except the magnetic field and scattering angles) are treated as free in the fit, though in the Bayesian least-squares minimization they are unlikely to deviate much outside the interval indicated by their prior value and uncertainty. These assumptions, i.e.\ nominal noise, free nuisance parameters, and $\alpha$-particles fitted by a slowing-down distribution, are likely the most realistic of the cases studied here.

Figure~\ref{fig,series3} shows representative example results for this case, again for a combined integration time of $\Delta t = 5 \times 20$~ms = 100~ms.
The results  are not substantially different from those of Series~1, but discrepancies with respect to the true values and in particular the associated uncertainties do show significant increases. 
Note that since the nuisance parameters related to thermal ions can now vary, so can the location of the automatically assigned velocity nodes outside the thermal bulk in the reconstructed $g(u)$. This means that the {\em true} observable fast-ion densities obtained within a specific projected velocity interval may vary between fits, so when plotting these we show their mean and standard deviation  in Figure~\ref{fig,series3}({\em b}) and subsequent figures.

\begin{figure}
\begin{center}
\includegraphics[width=8.2cm]{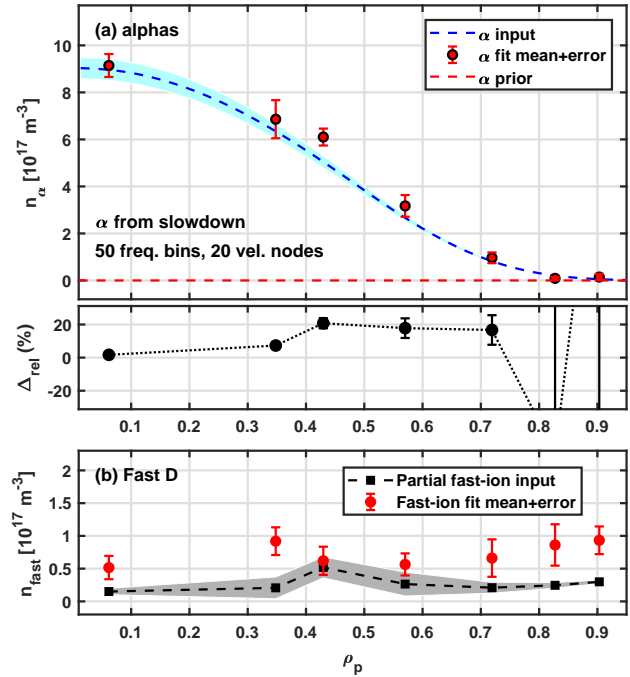}
\caption{Example results of fit series~2 with nominal noise and free nuisance parameters.}
\label{fig,series3}
\end{center}
\end{figure}

Since the synthetic spectra can now vary more significantly due to the increased noise levels, each density profile reconstructed from five spectra
for each receiver represents only an {\em example} outcome of a 100~ms measurement. Hence, to verify the generality of the results in Figure~\ref{fig,series3}, we repeated each $\Delta t=20$~ms fit series 100 times for each receiver. This further allows us to derive a  reliable mean accuracy and uncertainty on $n_\alpha$ as determined from each receiver, representing the predicted typical accuracy of the CTS diagnostic for 20~ms measurements. We list the resulting absolute and relative uncertainties obtained from this procedure in Table~\ref{tab,results}. For the five receiver views at $\rho_p  < 0.8$ (again corresponding to $n_\alpha \gtrsim 1\times 10^{17}$~m$^{-3}$), the values in the Table imply a mean accuracy on $n_\alpha$ as defined in equation~(\ref{eq,drel}) of $\Delta^*_{\rm rel} = 0.11 \pm 0.02$. 

We thus conclude that the more realistic assumptions adopted here compared to those of fit series~1 only lead to a marginal degradation in the accuracy of the reconstructed $\alpha$-particle density profile. Moreover, $n_\alpha$ can be constrained to significantly better than 20\% with a time resolution of 20~ms, well within the ITER measurement requirements above the relevant density limit. The fitted fast deuterium densities outside the thermal bulk also remain comparable to their true values but are now subject to significantly larger uncertainties.

\begin{table}
\caption{\label{tab,results} Typical absolute and relative accuracies, $\Delta_{\rm abs} = \langle n_{\alpha,{\rm fit}} -n_{\alpha,{\rm true}} \rangle$ and $\Delta_{\rm rel} =  \Delta_{\rm abs}/n_{\alpha,{\rm true}}$ for each receiver, in order of increasing central $\rho_p$ of the measurement volume. Values for each receiver are based on 100 fits in fit series~2, each corresponding to a time resolution of $\Delta t = 20$~ms.}
\begin{indented}
\item[]\begin{tabular}{@{}*{4}{ccccr}}
\br          
Receiver & $\rho_p$ & $n_{\alpha,{\rm true}}$ & $\Delta_{\rm abs}$ & $\Delta_{\rm rel}$ \cr
 & &  ($10^{17}$~m$^{-3}$) & ($10^{17}$~m$^{-3}$) & (\%) \cr
\mr
2 & 0.06 & 8.99 &	$0.57$	& $6$  \cr  
3 & 0.35 & 6.39 &	$0.86$	& $13$  \cr   
1 & 0.43 & 5.05 &	$0.96$	& $19$  \cr  
4 & 0.57 & 2.69 &	$0.24$	& $9$ \cr  
5 & 0.72 & 0.83 &	$0.06$	& $7$   \cr  
6 & 0.83 & 0.21 &	$-0.04$	& $-21$   \cr  
7 & 0.90 & 0.06 &	$0.10$	&$172$  \cr  
\br
\end{tabular}
\end{indented}
\end{table}

\subsection{Fit series 3: Enhanced noise and free nuisance parameters }\label{sec,series4}

We now discuss two examples with free nuisance parameters (as in series~2) and significantly elevated ECE background. As the ITER CTS diagnostic will operate at 60~GHz, below the fundamental electron cyclotron resonance at the nominal ITER field of $B_t = 5.3$~T, evaluation of the anticipated diagnostic background is not straightforward and is associated with some uncertainty. To test the impact of a much higher ECE background, which could potentially be relevant in high--$T_e$ ($\gtrsim 30$~keV) plasma scenarios \cite{rasm19},
we have repeated fit series~2 assuming $P_b = 2$~keV, but with all other parameters unaltered. In the case of 50~frequency channels, this raises the typical statistical noise level from 0.5~eV (our adopted noise floor) to 2.0~eV per channel, and the ECE background then clearly dominates the noise at most frequencies. 

The resulting averaged ($\Delta t = 100$~ms) profiles in Figure~\ref{fig,series4} suggest that the $\alpha$-particle densities remain reasonably well recovered, but there is a tendency for these to be overestimated at all radii. As previously, the robustness of this systematic over-estimation
was verified by running 100 fit series with $\Delta t = 20$~ms, in order to obtain average accuracies on the reconstructed $\alpha$-particle profiles. Figure~\ref{fig,series4_2} illustrates the outcome of averaging these individual fit series. This confirms that
the overall shape of the true $\alpha$-particle profile remains fairly well matched, but the reconstructed profile displays an offset of $0.6\pm 0.1 \times 10^{17}$~m$^{-3}$ averaged across all receivers. 

\begin{figure}
\begin{center}
\includegraphics[width=8.2cm]{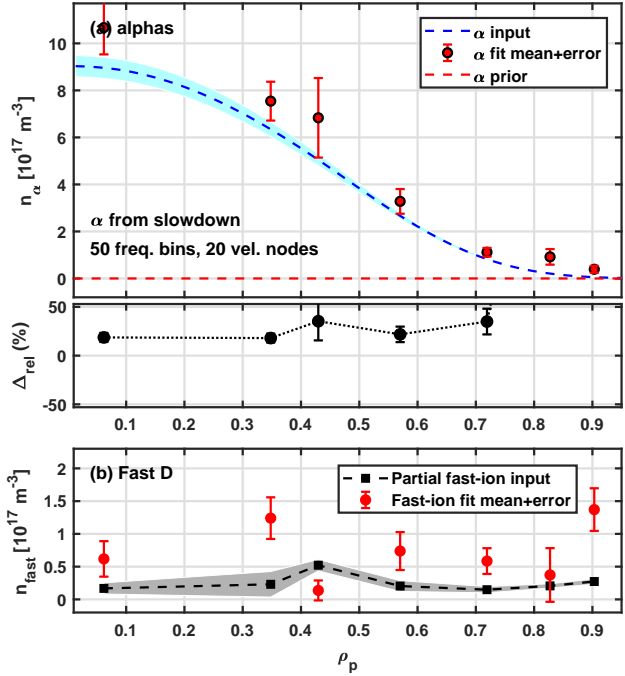}
\caption{Example results of fit series~3, with increased ECE background ($P_b=2$~keV) and free nuisance parameters.}
\label{fig,series4}
\end{center}
\end{figure}

\begin{figure}
\begin{center}
\includegraphics[width=8.2cm]{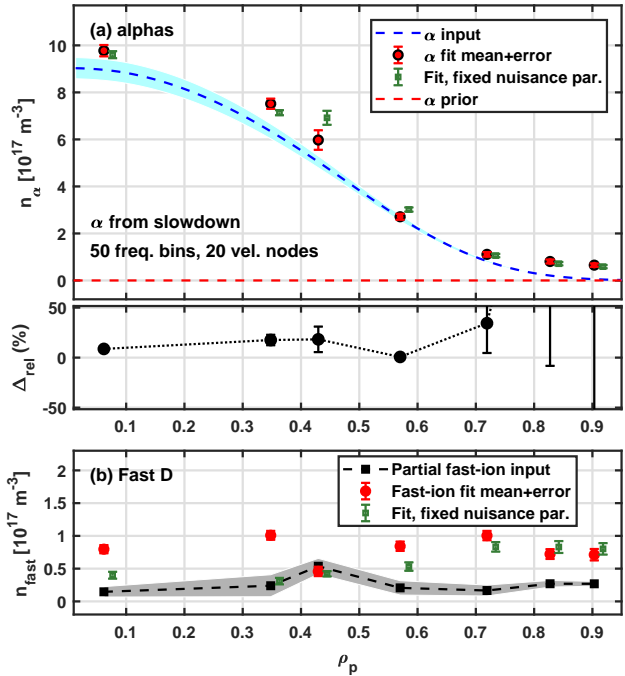}
\caption{As Figure~\ref{fig,series4}, but showing the average of 20 such profile reconstructions. Also shown are the results of fixing nuisance parameters in the fits, slightly displaced along the x-axis for ease of comparison.}
\label{fig,series4_2}
\end{center}
\end{figure}

Forcing non-negativity of $g(u)$ at all $u$, in the expectation that it could raise the fitted fast-D densities and correspondingly lower the $\alpha$-particle densities, did not significantly improve the $\alpha$-particle density recovery. Nor do we find the offset to be associated with a systematic under- or overestimation of one or more nuisance parameters. As demonstrated in Figure~\ref{fig,nuisance}, all fit results for nuisance parameters are symmetrically distributed around their true values. Furthermore, Figure~\ref{fig,series4_2} shows that fixing the nuisance parameters does not significantly improve the accuracy of recovered $\alpha$-particle densities, suggesting that the offset is dominated by the impact of increased noise in the spectra. 

We also note from Figure~\ref{fig,series4_2} that a similar systematic offset is seen for the fitted fast-D densities,
indicating that a strongly elevated background leads to an overestimate of the density of {\em all} fast-ion species with the present approach. 
However, the reconstruction of the observable fast-D densities does improve when fixing all thermal-ion parameters. Further work will be necessary to pinpoint the exact origin of the offset in $\alpha$-particle densities, establish its generality across plasma scenarios, and identify measures to improve the reconstruction of the absolute $\alpha$-particle profile under the adopted assumptions.

\begin{figure}
\begin{center}
\includegraphics[width=8.2cm]{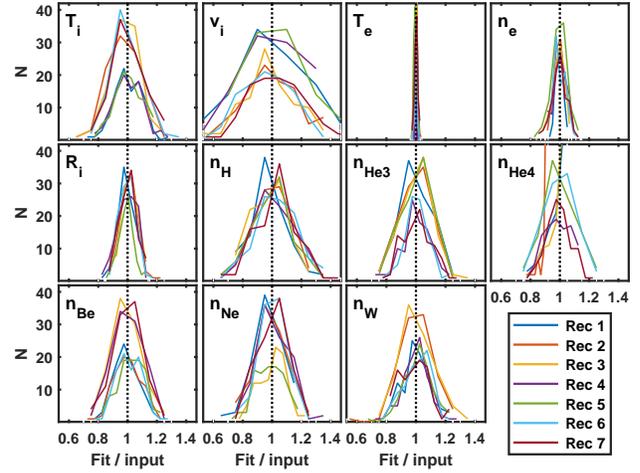}
\caption{Histograms of the ratio of fitted to true values for all free nuisance parameters in the 100 fit series shown in Figure~\ref{fig,series4_2}.}
\label{fig,nuisance}
\end{center}
\end{figure}

Nevertheless, even with this offset, the true and fitted absolute $\alpha$-particle densities are, encouragingly, in reasonable agreement.
For the five views at $\rho_p < 0.8$, the results based on all 100 fit series in Figure~\ref{fig,series4_2} with free nuisance parameters imply that $n_\alpha$ is recovered to within $\Delta^*_{\rm rel} = 0.16 \pm 0.06$\ on average, again consistent with the ITER measurement requirements on $n_\alpha$.
This suggests that a significantly increased ECE background of $\sim 2$~keV may diminish, but not compromise, the ability to extract reliable $\alpha$-particle profile measurements at 20~ms time resolution in the ITER baseline scenario.

\subsection{Fit series 4: Nominal noise, free nuisance parameters, and no fast-ion model assumptions}\label{sec,series5}

Our final case highlights the effects of fitting the spectra, which include both $\alpha$-- and NBI-particles, with a model including only a single, non-parameterized fast-ion velocity distribution $g(u)$, for which the fast-ion mass and charge are fixed at those of $\alpha$-particles. This approach leaves out the parameterized classical slowing-down distribution used to model the $\alpha$-particles in series 1--3, and thus makes no assumptions about the shape of the fast-ion velocity distribution. A representative result of such fits is shown in Figure~\ref{fig,series5}, in the form of fast-ion fit results representing velocity integrals of the fitted $g(u)$ outside the thermal bulk. These are compared to 
the sum of true observable fast-ion densities ($\alpha$ + NBI) computed in a similar manner, along with the true total $\alpha$ and fast-ion ($\alpha$ + NBI) densities.
The fitted observable "densities" should here not be interpreted as physical fast-ion densities but rather as the equivalent density of $\alpha$-particles producing the fast-ion signal on their own, so caution should be exercised when interpreting the results.

\begin{figure}
\begin{center}
\includegraphics[width=8.2cm]{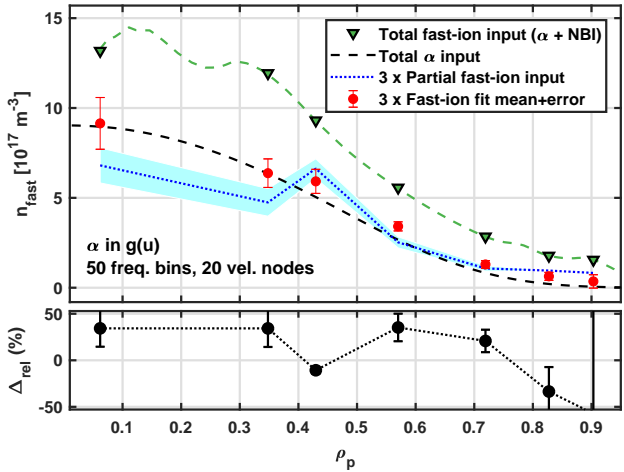}
\caption{Example of true and fitted total and observable fast-ion densities for fit series~4. The observable fast-ion values represent the velocity-integrated sum of $\alpha$-particle and NBI densities at projected velocities outside the thermal bulk and are scaled by a factor of 3 for ease of comparison to the true total densities.}
\label{fig,series5}
\end{center}
\end{figure}

The fitted  observable values, based on allowing for $\alpha$-particles only, are generally comparable to the corresponding true values representing
the sum of $\alpha$ + NBI densities. 
However, the shape of the inferred profile is more similar to that of the true {\em total} $\alpha$-particle density in this case, despite the lack of a straightforward correspondence between the two quantities. Most notably, the fits slightly over- and underestimate the true values for the inner- and outermost receivers, respectively, and they do not clearly reproduce the kink in the true $\alpha$ + NBI profile for Receiver~1 at $\rho_p=0.42$. These results are generic, as found from repeating this fit series 100~times.

Analogous to previous fit series, we compute for each receiver the quantity $\langle |n_{\rm fast,fit}- n_{\rm fast,true}|\rangle/ \langle n_{\rm fast,true}\rangle$, where all values are now observable, i.e.\ partial, ones and the true values are included in the averaging since they can vary. Averaging the results across the innermost five receivers yields $\Delta^*_{\rm rel} =0.22\pm 0.03$\, which can be compared to $\Delta^*_{\rm rel} \approx 0.1$ obtained for fit series 1--3. Results of this case hence show that a model with no assumptions regarding the functional form of $g(u)$ does recover the true densities reasonably well, but 
the accuracy is somewhat poorer, the overall shape of the inferred radial profile is not reproduced in detail, and, importantly, the fit result assumes a fast-ion contribution from $\alpha$-particles only as we cannot distinguish between the contribution from $\alpha$-- and NBI particles. This is a consequence of using a single $g(u)$ to represent  a degenerate combination of two fast-ion populations with different ion mass and charge but overlapping velocity ranges. In the next section, we discuss possible options for improving on this.

\section{Discussion}\label{sec,discuss}

\subsection{Addressing the ITER measurement requirements}

Formally, ITER diagnostics must be able to measure the confined $\alpha$-particle density profile down to $n_\alpha = 10^{17}$~m$^{-3}$ at an accuracy of 20\%, with a temporal  resolution of $\Delta t=100$~ms and a spatial resolution of 10--100~cm for CTS specifically and $a/10$ for the combined ITER suite of diagnostics, where $a=2.0$~m is the ITER minor radius \cite{donn07,idm1,iter}.
A detailed discussion of the spatial resolution of the CTS diagnostic will be presented elsewhere. For now, we note that the maximum extent of the scattering volumes as defined in Figure~\ref{fig,raytracing} range from 12--80~cm in the adopted plasma scenario, thus conforming to the above requirements.
Our results hence suggest that the CTS diagnostic will be able to meet, and in some cases surpass, the measurement requirements in the ITER baseline plasma scenario. To the extent that the $\alpha$-particle dynamics in ITER may be characterized by a classical slowing-down distribution, Section~\ref{sec,series3} shows that $n_\alpha$ can be recovered from CTS spectra to within $\approx 10$\% of the true value above $10^{17}$~m$^{-3}$ at 20~ms time resolution. 

This applies under conservative assumptions about the ECE background and under measurement conditions in which all relevant nuisance parameters are fitted along with the fast-ion contribution. Under more favourable but not necessarily unrealistic conditions, in which the ECE background at 55--65~GHz is negligible and thermal plasma parameters can be well constrained from other measurements, the CTS diagnostic can perform even better. As demonstrated in Section~\ref{sec,series1}, an absolute accuracy on the recovered  $\alpha$-particle densities of $\sim 0.1\times 10^{17}$~m$^{-3}$ for all seven CTS receivers may be achieved in this case, with a relative accuracy of just 7\%  at $n_\alpha > 1\times 10^{17}$~m$^{-3}$ for an integration time of 20~ms. 

This comes about despite (1) the fact that the spectral distortions induced by refraction are not taken into account in the model spectra that are fitted to the data (a computationally expensive inverse problem which is still open); (2) the presence of multiple fast-ion populations in the plasma with partially overlapping velocity ranges; (3) 
the use of a model distribution function for the $\alpha$-particles that is not identical to that underlying the synthetic data; (4) the presence of magnetosonic peaks in the spectra that mask the $\alpha$-particle contribution across part of the relevant frequency range; 
and (5) the fact that any CTS diagnostic is sensitive to ions across only across a limited range in pitch $p= v_\parallel/v$ \cite{sale11}, where $v$ is the fast-ion velocity and $v_\parallel$ its component along the magnetic field, the exact range generally depending on ion energy and projection angle $\phi$ and being $|p| \lesssim 0.9$ for the values of $\phi=96^\circ-100^\circ$ relevant here. We thus conclude that these effects do not significantly hamper the recovery of the $\alpha$-particle density using existing fitting algorithms, nor the inferred partitioning of fast ions into $\alpha$-particles and NBI ions across the plasma minor radius.

Even when assuming a functional form of the $\alpha$-particle distribution function, simultaneous recovery of the density of other fast-ion species across parts or all of velocity space in the presence of a significant $\alpha$-particle population may be less straightforward with existing inversion algorithms. The absolute discrepancies in fast-D densities with respect to the true (observable) values are comparable to those pertaining to the recovered $\alpha$-particle density profile, but the relative differences are significantly larger. Optimization of fit parameters and algorithmic choices may improve on this, but an exhaustive exploration of this is beyond our present scope.

If the $\alpha$-particles must be treated using a fully free non-parameterized distribution function, a degeneracy results from the simultaneous presence of multiple fast-ion species in the plasma. In this case, not even a combined fast-ion density (e.g., $\alpha$ + NBI) can be derived from CTS spectra, only an equivalent density that would result from assuming a single fast-ion species. The result is not straightforward to interpret, except in the trivial case where one species is expected to clearly dominate over others.
Furthermore, the result will be an {\em observable} value integrated over some projected velocity interval, rather than the true total density, since fast ions are masked by the thermal bulk if their velocity projected along ${\bf k}^\delta$ is small -- however large in 3D. Under this condition, existing fitting algorithms cannot provide an unambiguous total $\alpha$-particle density from CTS spectra alone, but we discuss possible steps to improve this below.

We emphasize that these results apply to the (adopted version of) the ITER baseline plasma scenario only.  At fixed $T_e$, characterization of $f_\alpha$ with CTS is favoured by such high-density scenarios, since the fusion rate scales more strongly with $n_e$ than the opposing effect of reducing the $\alpha$-particle slowing-down time $\tau_{sd}$ (e.g., \cite{eged05}). Planned operating scenarios with lower $n_e$ and higher $T_e$, such as the $I_p= 12.5$~MA hybrid and $I_p=9$~MA steady-state scenarios, should be subject to increased NBI penetration but also longer $\tau_{sd}$ and comparable $\alpha$-particle production. In future work, we aim to extend the present study to include these and other plasma scenarios.

\subsection{Limitations of the fitting procedure}

Our approach is based on using existing fitting algorithms to extract $\alpha$-particle densities from synthetic CTS spectra. Here we will briefly discuss possible limitations associated with this.

First, our Bayesian fitting algorithm, based on comparing forward-modelled spectra to measurements, enables us to account for the contribution from the many parameters that influence CTS spectra, cf.\ Table~\ref{tab,nuisance}, along with their uncertainties as quantified by other measurements. However, for the chosen number of frequency bins in the synthetic spectra, the fitting algorithm might not always converge on the global minimum of our generalized $\chi^2$ goodness-of-fit parameter in this multi-dimensional parameter space \cite{stej12}. Hence, there could potentially be inaccuracies associated with the fitting procedure itself, but these are non-trivial to identify, let alone quantify, without benchmarking against other algorithms which is beyond our scope here.

We have also assumed a specific number of frequency channels in the spectra and number of velocity nodes in the reconstructed $g(u)$. This was done to mimic typical filter bank setups for fast-ion measurements in existing CTS implementations, for which our fitting algorithm was developed. As such, our numerical setup is oriented towards recovery of the fast-ion information contained in the spectral wings at large frequency shifts from $\nu_{\rm gyr}$ and so does not make full use of the information contained in the thermal features of the spectra. Preliminary tests with increasing the spectral resolution of the spectra and/or the velocity resolution of $g(u)$ to better account for those features do indicate improved accuracy on the recovered $\alpha$-particle and fast-D densities at our nominal noise level (fit series~2), but not so in the case of enhanced noise (series~3), and any improvements are at the expense of heavier computational load.
Development of efficient algorithms to self-consistently fit  both thermal and fast-ion features in the spectra could strengthen constraints on the fast-ion components, especially in cases of  high noise levels or nuisance parameters that are not well constrained by other diagnostics. 

Another improvement would be to extend the fitted model to account for effects of refraction, e.g.\ by allowing the Bayesian priors for nuisance parameters to include information on the typical magnitude of such effects and implementing the methods described in Section~\ref{sec,step2} when calculating the fitted spectra. In fact, the frequency dependence of refraction imples that the measurement volume of one receiver at 55~GHz may overlap with that of another receiver at 65~GHz, and it might be possible to take advantage of this in data analysis using tomographic techniques.

A demonstration of how features of the fitting algorithm can impact some of our results is provided by the kink seen for Receiver~1 at  $\rho_p \approx 0.4$ in the true  observable fast-ion profiles of Section~\ref{sec,results}. 
With its scattering volume  located on the high-field side,  Receiver~1 has a scattering geometry with a larger $\theta =\angle({\mathbf k}^i,{\mathbf k}^s)$
and hence longer ${\bf k}^\delta = {\bf k}^s -  {\bf k}^i$ than the receivers adjacent to it in poloidal flux space. As our
fitting algorithm places velocity nodes at $u_n \approx 2\pi (\nu_n - \nu_{\rm gyr})/k^\delta$, where $\nu_n$ is the center frequency of the $n$'th spectral bin, nodes will be distributed across a smaller interval in $u$ when $k^\delta$ is larger, given a fixed frequency range of the spectra. With the innermost velocity nodes in $g(u)$ thus placed closer to the thermal bulk, where $g(u)$ increases steeply, the resulting true {\em observable} fast-ion density outside the thermal bulk will be larger at comparable {\em total} density.
Fixing both the location and number of velocity nodes outside the thermal bulk would thus remove the resulting kink.

A different issue is our choice of model distribution function $f_\alpha$ for the $\alpha$-particles when fitting the spectra. Here we have shown that adopting a classical slowing-down distribution function in the fits performs well in terms of describing the total ion density implied by our simulated $f_\alpha$. However, the true $f_\alpha$ in ITER will differ somewhat from a velocity-isotropic slowing-down distribution even in case of neoclassical transport \cite{yako15}. The differences would be exacerbated by $\alpha$-particles interacting with Alfv\'en eigenmodes and other magnetohydrodynamic fluctuations \cite{pinc15}, or even with microturbulence in the plasma \cite{estr06}. The impact of this could be studied using different synthetic $f_\alpha$ or by introducing physically motivated distortions to the distribution function used here.

The above issue is circumvented if fitting all fast ions in the plasma by a free, non-parameterized $g(u)$ as in Section~\ref{sec,series5}. In that case, however, it might be beneficial to apply a Bayesian prior on the contribution from ions accelerated by NBI or ion cyclotron resonance heating. Such a prior could potentially be characterized experimentally in collisionality-matched pure D discharges with negligible $\alpha$-particle contribution, allowing improved constraints on $n_\alpha$ in D-T plasmas. In a similar vein, prior information (or an improved fitting model) for $\alpha$-particles might be obtained experimentally in D-T plasmas during short periods with no direct ion heating, using electron heating to maintain the plasma collisionality and stored energy. 

More fundamentally, the results of Section~\ref{sec,series5} highlight the relevance of developing novel analysis algorithms that can treat multiple fast-ion populations simultaneously. A simple solution would be to allow two or more non-parameterized $g(u)$ in the fit, each representing ion populations with specified mass, charge, and 3D velocity ranges, and apply an informed fit prior on their density ratio at each projected velocity from transport simulations and/or measurements by other diagnostics. In addition, improved constraints on $f_\alpha$ could be derived from combining CTS measurements with other diagnostic input using fast-ion velocity-space tomography\cite{sale18}.

\section{Conclusions and outlook}\label{sec,concl}

We have developed and applied a framework for evaluating the performance of the planned ITER CTS diagnostic in terms of its ability to recover fast-ion densities, notably those of confined $\alpha$-particles, which is its primary purpose. The procedure is aimed at generating fully realistic ITER CTS spectra and replicating the analysis that could be applied to actual ITER CTS measurements. Our approach is generic, however, in the sense that it can be used to evaluate the performance of a potential CTS diagnostic at any device and operating frequency, provided reliable estimates of plasma  and fast-ion parameters.

The present analysis is the first to apply the actual CTS diagnostic design to be implemented in ITER, and the first to combine the effects of noise, frequency-dependent refraction, and the presence of multiple fast-ion populations in the computation of synthetic ITER CTS spectra. It is also the first to invert such spectra to recover information on the underlying fast-ion populations, and hence the first to provide a realistic assessment of  how well the ITER CTS diagnostic will meet the ITER measurement requirements on confined fusion-born $\alpha$-particles. 

As part of this work, we have constructed a CTS model that accounts for the impact of frequency-dependent refraction on modelled CTS spectra. Applying this to the high-density ITER baseline plasma scenario, we find that this effect is generally small, inducing spectral distortions of typically $\lesssim  0.5$~eV for most frequencies and receivers. Encouragingly, fitting a model to synthetic ITER CTS data that does {\em not} take this effect into account suggests that frequency-dependent refraction will not significantly affect the interpretation of the measurements. Other D-T plasma scenarios are expected to induce less refraction at the relevant frequencies due to the lower electron density and higher temperature.

For an assumed ECE background of 100~eV and if adopting a classical slowing-down distribution for the $\alpha$-particles when fitting the spectra, the resulting radial profiles of total $\alpha$-particle density can be constrained to within $\sim 10$\% in regions where it is non-negligible (cf.\ Table~\ref{tab,results}). This assumes a time resolution (integration time) of 20~ms, thus easily meeting the ITER CTS measurement requirements on the accuracy and temporal resolution of  
the recovered $\alpha$-particle density profiles of 20\% and 100~ms, respectively. Even if assuming a high ECE background of 2~keV, the overall shape of the $\alpha$-particle density profile can still be well recovered. While there is then a tendency to systematically overestimate $n_\alpha$, the true densities are still reconstructed to within  $\sim 20$\% for all receivers probing regions of $n_\alpha \gtrsim 1 \times 10^{17}$~m$^{-3}$, again consistent with the ITER measurement requirements.

These encouraging results arise despite a number of elements in the analysis that are potentially detrimental to accurate recovery of fast-ion densities. These include ignoring frequency-varying refraction when interpreting the spectra, applying a model distribution function for the $\alpha$-particles that does not match the true one, and the degeneracy resulting from the presence of multiple fast-ion populations in the plasma with partially overlapping velocity ranges. Thus, we also conclude that these effects alone do not significantly hamper the inversion of CTS measurements under ITER--relevant conditions using existing algorithms. With respect to the concern that the presence of beam ions could mask the $\alpha$-particle signal in ITER CTS spectra, our conclusion extends earlier results \cite{eged05,sale09a,sale09b} to encompass the recovery of the respective fast-ion densities from measurements.

Summarizing, the present design of the ITER CTS diagnostic should be able to meet the ITER measurement requirements
 regarding the accuracy and time resolution with which the density profile of fusion $\alpha$-particles can be inferred in the baseline plasma scenario. Tailoring of our inversion algorithm or development of entirely new ones, for example based on machine learning \cite{vand18}, might further improve the accuracy of the reconstructed fast-ion densities. Based on the approach adopted here, further studies are planned for assessing the generality of our results across a wider range of ITER plasma scenarios.

\section*{Acknowledgments}

This article is dedicated to the memory of co-author and colleague Frank Leipold, who passed away during its preparation.

We thank F.~Koechl and A.~Polevoi for help with providing ITER plasma scenario data. The work leading to this publication has been funded partially by
Fusion for Energy under Grant F4E-FPA-393. This publication reflects the views only of the authors, and Fusion for Energy cannot be held responsible for any use which may be made of the information contained therein.


\section*{References}
\begin{thebibliography}{99}

\bibitem{chug11} Chugunov I.N. {\em et al} 2011 {\em Nucl. Fusion} {\bf 51} 083010

\bibitem{noce17} Nocente M. {\em et al} 2017 {\em Nucl. Fusion} {\bf 57} 076016

\bibitem{sale09b} Salewski M. {\em et al} 2009 {\em Plasma Phys. Control. Fusion} {\bf 51} 035006

\bibitem{kors16} Korsholm S.B. {\em et al} 2016 High power microwave diagnostic for the fusion energy experiment ITER {\em Conf.\ Proc.\ for 41st Int.\ Conf.\ on Infrared, Millimeter and Terahertz Waves (IRMMW-THz)} (https://ieeexplore.ieee.org/abstract/document/7758537/)

\bibitem{kors19} Korsholm S.B. {\em et al} 2019  {\em EPJ Web Conf.} {\bf 203} 03002

\bibitem{sale15} Salewski M. {\em et al} 2015 {\em Nucl. Fusion} {\bf 55} 093029

\bibitem{sale18} Salewski M. {\em et al} 2018 {\em Nucl. Fusion} {\bf 58} 096019

\bibitem{kipt90} Kiptily V.G. 1990 {\em Fusion Sci. Technol.} {\bf 18} 583

\bibitem{walt14} Waltz R.E. and Bass E.M. 2014 {\em Nucl. Fusion} {\bf 54} 104006

\bibitem{donn07} Donn\'e A.J.H. {\em et al} 2007 {\em Nucl. Fusion} {\bf 47}  S337

\bibitem{idm1} Vayakis G. and Philip A. 2014 Diagnostic Measurement requirements with relation to diagnostic systems, ITER IDM document ID ITER\_D\_2LMA3T v1.0 {\em Technical Report} https://user.iter.org/?uid=2LMA3T

\bibitem{iter} ITER diagnostics Database v.~1.18,  ITER IDM document ID ITER\_D\_26L94U  https://user.iter.org/?uid=26L94U

\bibitem{rasm19}  Rasmussen J. {\em et al} Modelling the electron cyclotron emission below the fundamental resonance in ITER, {\em Plasma Phys. Control. Fusion} in press https://iopscience.iop.org/article/10.1088/1361-6587/ab2f4a

\bibitem{hutc85} Hutchinson D.P. {\em et al} 1985 {\em Rev. Sci. Instrum.} {\bf 56} 1075

\bibitem{orsi97} Orsitto F. and Giruzzi G. 1997  {\em Rev. Sci. Instrum.} {\bf 68} 686

\bibitem{bind03} Bindslev H., Meo F. and Korsholm S.B. 2003 ITER Fast Ion Collective Thomson Scattering Feasibility study, {\em Technical Report} Ris{\o} National Laboratory, Ris\o, Denmark  http://orbit.dtu.dk/en/publications/iter-fast-ion-collective-thomson-scattering(401e562c-65a3-4345-9032-da5cc8b9013b).html

\bibitem{bind04} Bindslev H. {\em et al} 2004 {\em Rev. Sci. Instrum.} {\bf 75} 3598

\bibitem{meo04} Meo F. {\em et al} 2004 {\em Rev. Sci. Instrum.} {\bf 75} 3585

\bibitem{kors06} Korsholm S.B. {\em et al} 2006 {\em Rev. Sci. Instrum.} {\bf 77} 10E514

\bibitem{meo07} Meo F., Bindslev H. and Korsholm S.B. 2007 ITER fast ion collective Thomson scattering, conceptual design of 60 GHz system, Ris\o-R-1600(EN) {\em Technical Report} http://www.orbit.dtu.dk/en/publications/iter-fast-ion-collective-thomson-scattering-conceptual-design-of-60-ghz-system(30e4aab5-22c1-43e0-91a1-39f12584b782).html
 
\bibitem{tsak08} Tsakadze E. {\em et al} 2008 {\em Fusion Sci. Technol.} {\bf 53} 69

\bibitem{kors08} Korsholm S.B. {\em et al} 2008 {\em AIP Conf. Proc.} {\bf  988}, 118

\bibitem{mich09a} Michelsen S. {\em et al} 2009 ITER Fast Ion Collective Scattering, development of diagnostic components and techniques, Ris\o-R-1716(EN) {\em Technical Report}  http://orbit.dtu.dk/en/publications/efda-task-tw6tpdsdiadev-deliverable-2-iter-fast-ion-collective-scattering-development-of-diagnostic-components-and-techniques(77b19428-9759-4ce7-84af-0ea6f093b37e).html

\bibitem{mich09b} Michelsen P.K. {\em et al} 2009 Engineering design of the ITER Collective Thomson Scattering diagnostic, Ris\o-R-1717(EN) {\em Technical Report}
http://orbit.dtu.dk/en/publications/engineering-design-of-the-iter-collective-thomson-scattering-diagnostic-contract-efda-061478(5d9b474b-b184-4da2-be9c-eb59c0558fe0).html

\bibitem{bind05} Bindslev H. {\em et al} 2005 ITER Fast Ion Collective Thomson Scattering, detailed integrated design of the collective Thomson scattering (CTS) system for ITER {\em Technical Report} http://orbit.dtu.dk/en/publications/iter-fast-ion-collective-thomson-scattering(220dc911-ddcb-43a5-9f96-c7a2a0153f86).html

\bibitem{kors10} Korsholm S.B. {\em et al} 2010  {\em Rev. Sci. Instrum.} {\bf 81} 10D323

\bibitem{stej12} Stejner M. {\em et al} 2012 {\em Nucl. Fusion} {\bf 52} 023011 

\bibitem{eged05} Egedal J. {\em et al} 2005 {\em Nucl. Fusion} {\bf 45} 191    

\bibitem{sale09a} Salewski M. {\em et al} 2009  {\em Nucl. Fusion} {\bf 49} 025006

\bibitem{sale11} Salewski M. {\em et al} 2011 {\em Nucl. Fusion} {\bf 51} 083014

\bibitem{infa17} Infante V.  {\em et al} 2017 {\em Fusion Eng. Des.} {\bf 123} 663

\bibitem{lope18} Lopes A.  {\em et al} 2018 {\em Fusion Eng. Des.} {\bf 134} 22

\bibitem{vida19} Vidal C.  {\em et al} 2019 {\em Fusion Eng. Des.} {\bf 140} 123

\bibitem{bind99}  Bindslev H. 1999  {\em Rev. Sci. Instrum.} {\bf 70} 1093

\bibitem{niel08} Nielsen S. K. {\em et al} 2008 {\em Phys. Rev. E} {\bf 77} 016407 

\bibitem{sale10} Salewski M.  {\em et al} 2010 {\em Nucl. Fusion} {\bf 50} 035012

\bibitem{niel11} Nielsen S. K. {\em et al} 2011 {\em Nucl. Fusion} {\bf 51} 063014

\bibitem{rasm15}  Rasmussen J. {\em et al} 2015 {\em Plasma Phys. Control. Fusion} {\bf 57} 075014

\bibitem{niel17} Nielsen S.K.  {\em et al}  2017 {\em Phys. Scr.} {\bf 92} 024001

\bibitem{bind92a} Bindslev H. 1992 {\em Plasma Phys. Control. Fusion} {\bf 34} 1601

\bibitem{bind92b} Bindslev H. 1992 On the Theory of Thomson Scattering and  Reflectometry in a Relativistic Magnetized Plasma, {\em PhD Thesis} Ris{\o} National Laboratory, Roskilde, Denmark, Ris\o-R-663

\bibitem{bind93} Bindslev H. 1993 {\em Plasma Phys. Control. Fusion} {\bf 35} 1093

\bibitem{shka86} Shkarofsky I.P. 1986 {\em J. Plasma Phys.} {\bf 35} 319

\bibitem{bind91} Bindslev H. 1991 {\em Plasma Phys. Control. Fusion} {\bf 33} 1775

\bibitem{hend15} Henderson M.  {\em et al} 2015 {\em Phys. Plasmas} {\bf 22} 021808

\bibitem{iter18} ITER Organization 2018  ITER Research Plan within the Staged Approach,  ITR-18-003 {\em ITER Technical Report}
https://www.iter.org/technical-reports

\bibitem{scen1} Simulation data performed as part of ITER-F4E project GRT502, ITER IDM document ID ITER\_D\_SQ8LP5  https://user.iter.org/?uid=SQ8LP5

\bibitem{imbe15} Imbeaux F.  {\em et al} 2015 {\em Nucl. Fusion} {\bf 55} 123006

\bibitem{pinc18} Pinches S. {\em et al} 2018  Progress in the ITER Integrated Modelling Programme and the ITER Scenario Database {\em Preprint: 2018 IAEA Fusion Energy Conf.}
(Gandhinagar, India) [TH/P6-7] (https://nucleus.iaea.org/sites/fusionportal/Shared Documents/FEC 2018/fec2018-preprints/preprint0741.pdf)

\bibitem{schn12} Schneider P.A. 2012  Characterization and scaling of the tokamak edge transport barrier, {\em PhD Thesis} Ludwig--Maximilians--Universit\"at, Munich, Germany 

\bibitem{bind93b} Bindslev H. 1993 {\em Plasma Phys. Control. Fusion} {\bf 35} 1615

\bibitem{bind96} Bindslev H. 1996 {\em J. Atmos. Terr. Phys.} {\bf 58} 983

\bibitem{fast} Simulation data, ITER IDM document ID ITER\_D\_Q43JH7  https://user.iter.org/?uid=Q43JH7

\bibitem{stej17b} Stejner M. {\em et al} 2017 {\em Plasma Phys. Control. Fusion} {\bf 59} 075009

\bibitem{stej10} Stejner M. {\em et al} 2010 {\em Rev. Sci. Instrum.} {\bf 81} 10D515

\bibitem{stej18} Stejner M. 2018 55.C7 CTS: Performance Analysis Report, ITER IDM document ID ITER\_D\_UXVCNQ v1.0 {\em Technical Report}  https://user.iter.org/?uid=UXVCNQ

\bibitem{niel15} Nielsen S.K. {\em et al} 2015 {\em Plasma Phys. Control. Fusion} {\bf 57} 035009 

\bibitem{rasm16}  Rasmussen J. {\em et al} 2016 {\em Nucl. Fusion} {\bf 56} 112014

\bibitem{furt12}  Furtula V. {\em et al} 2012 {\em Rev. Sci. Instrum.} {\bf 83} 013507  

\bibitem{hugh88} Hughes T.P. and Smith S.R.P. 1988 {\em Nucl. Fusion} {\bf 28} 1451 

\bibitem{yako15} Yavorskij V.  {\em et al} 2015  {\em J. Fusion Energy} {\bf 34} 774

\bibitem{pinc15} Pinches S. {\em et al} 2015  {\em Phys. of Plasmas} {\bf 22} 021807

\bibitem{estr06} Estrada-Mila C., Candy J. and Waltz R. E. 2006 {\em Phys. of Plasmas} {\bf 13} 112303

\bibitem{vand18} van den Berg J. {\em et al} 2018  {\em Rev. Sci. Instrum.} {\bf 89} 083507


\end{thebibliography}
\end{document}